\RequirePackage{snapshot}

\documentclass[10pt, conference, letterpaper]{IEEEtran}
\pdfoutput=1

\usepackage{cite}
\usepackage{amsmath}
\usepackage{amsthm}
\usepackage{mathtools}
\usepackage{amsfonts}
\usepackage{array}
\usepackage[caption=false,font=footnotesize]{subfig}
\usepackage{pgfplots}
\usepackage{tikz}
\usepackage{csvsimple}
\usepackage{thm-restate}
\usetikzlibrary{chains}
\usetikzlibrary{positioning}

\usepackage{color}

\newcommand{\allClasses}{K}
\newcommand{\class}{k}
\newcommand{\flow}{f}
\newcommand{\classRoutePair}{\flow}
\newcommand{\classRate}{x_\class}
\newcommand{\routeRate}{x_\classRoutePair}
\newcommand{\targetRouteRate}{A_\classRoutePair}
\newcommand{\transportUtility}{V_\classRoutePair(\routeRate)}

\title{MON: Mission-optimized Overlay Networks}
\author{
  \IEEEauthorblockN{B. Spang, A. Sabnis, R. Sitaraman, D. Towsley}
  \IEEEauthorblockA{College of Information \& Computer Sciences\\
  U. Massachusetts - Amherst\\
  Amherst, MA 01003\\
 \tt{ \{bspang,asabnis,ramesh,towsley\}@cs.umass.edu}}
\and
  \IEEEauthorblockN{B. DeCleene}
  \IEEEauthorblockA{BAE Systems \&  Technology Solutions\\ Burlington, MA 01803\\ \tt{brian.decleene@baesystems.com}}
}

\begin{document}

% \AddToShipoutPictureBG*{%
%   \AtPageLowerLeft{%
%     \setlength\unitlength{1in}%
%     \hspace*{\dimexpr0.5\paperwidth\relax}%%  change \dimexpr0.5\paperwidth\relax appropriately
%     \makebox(0,0.75)[c]{Approved for public release; unlimited distribution}%
% }}

\maketitle

\begin{abstract}
Large organizations often have users in multiple sites which are connected over the Internet. Since resources are limited, communication between these sites needs to be carefully orchestrated for the most benefit to the organization. We present a Mission-optimized Overlay Network (MON), a hybrid overlay network architecture for maximizing utility to the organization. We combine an offline and an online system to solve non-concave utility maximization problems. The offline tier, the Predictive Flow Optimizer (PFO), creates plans for routing traffic using a model of network conditions. The online tier, MONtra, is aware of the precise local network conditions and is able to react quickly to problems within the network. Either tier alone is insufficient. The PFO may take too long to react to network changes. MONtra only has local information and cannot optimize non-concave mission utilities. However, by combining the two systems, MON is robust and achieves near-optimal utility under a wide range of network conditions. While best-effort overlay networks are well studied, our work is the first to design overlays that are optimized for mission utility.
\end{abstract}
%%% Local Variables:
%%% mode: plain-tex
%%% TeX-master: "paper.tex"
%%% End:

\section{Introduction}

Large organizations have users in multiple sites that are connected over the Internet. A business may have multiple offices around the world which need to communicate with each other. A defense organization may have personnel deployed at multiple sites, who need to communicate and fulfill specific mission goals. A retailer may have multiple shops, warehouse locations, and offices. One traditional approach to facilitating communication between distributed sites of an organization is to deploy a {\em private} enterprise network with dedicated infrastructure to fulfill the organization's communication requirements. However, an alternate approach is to build an {\em overlay} on top of the {\em public} Internet, avoiding the need for dedicated infrastructure.

Overlays have been studied and built for the past 25 years \cite{Eriksson94, AndersenBalakrishnanKaashoekEtAl2001,StoicaMorrisLiben-NowellEtAl2003, RatnasamyFrancisHandleyEtAl2001,ZhaoKubiatowiczJoseph2002}. Large CDNs such as Akamai \cite{akamai-overview} have built overlays for delivering web and video content since the late 1990's\cite{sitaraman2014overlay}.  These overlays are ``best-effort'', in that they are concerned with providing higher reliability and performance than what the native Internet can offer for all traffic using the overlay.  Such best-effort overlays include  caching overlays for Web content \cite{dilley2002globally}, routing overlays for reliably transporting live video streams \cite{andreev2003designing,KontothanasisSWHKMSS04}, P2P overlays for downloads \cite{StoicaMorrisLiben-NowellEtAl2003, RatnasamyFrancisHandleyEtAl2001,ZhaoKubiatowiczJoseph2002}, and security overlays for preventing DDoS attacks \cite{sitaraman2014overlay}. However, best-effort overlays do not explicitly optimize the ``mission goals'' of the organization that operates the overlay.

In this paper, in contrast to best-effort overlays studied in prior work, we propose and study overlays that are driven by explicitly stated mission goals. For instance, consider a multi-site defense organization. The goals for the overlay are set by an operator who dictates the relative utility of various types of communication that occur between the different sites. Note that the mission goals may vary with time, e.g., an urgent all-hands video conference watched by users in all the sites may take higher precedence than downloads, VOIP and other traffic classes. A Mission-optimized Overlay Network (MON) dynamically allocates available overlay resources to the traffic between sites to maximize overall mission utility, enabling the goals of the organization to be met.

\subsection{MON Functionality}
\begin{figure}
  \centering
  \includegraphics[width=0.45\textwidth]{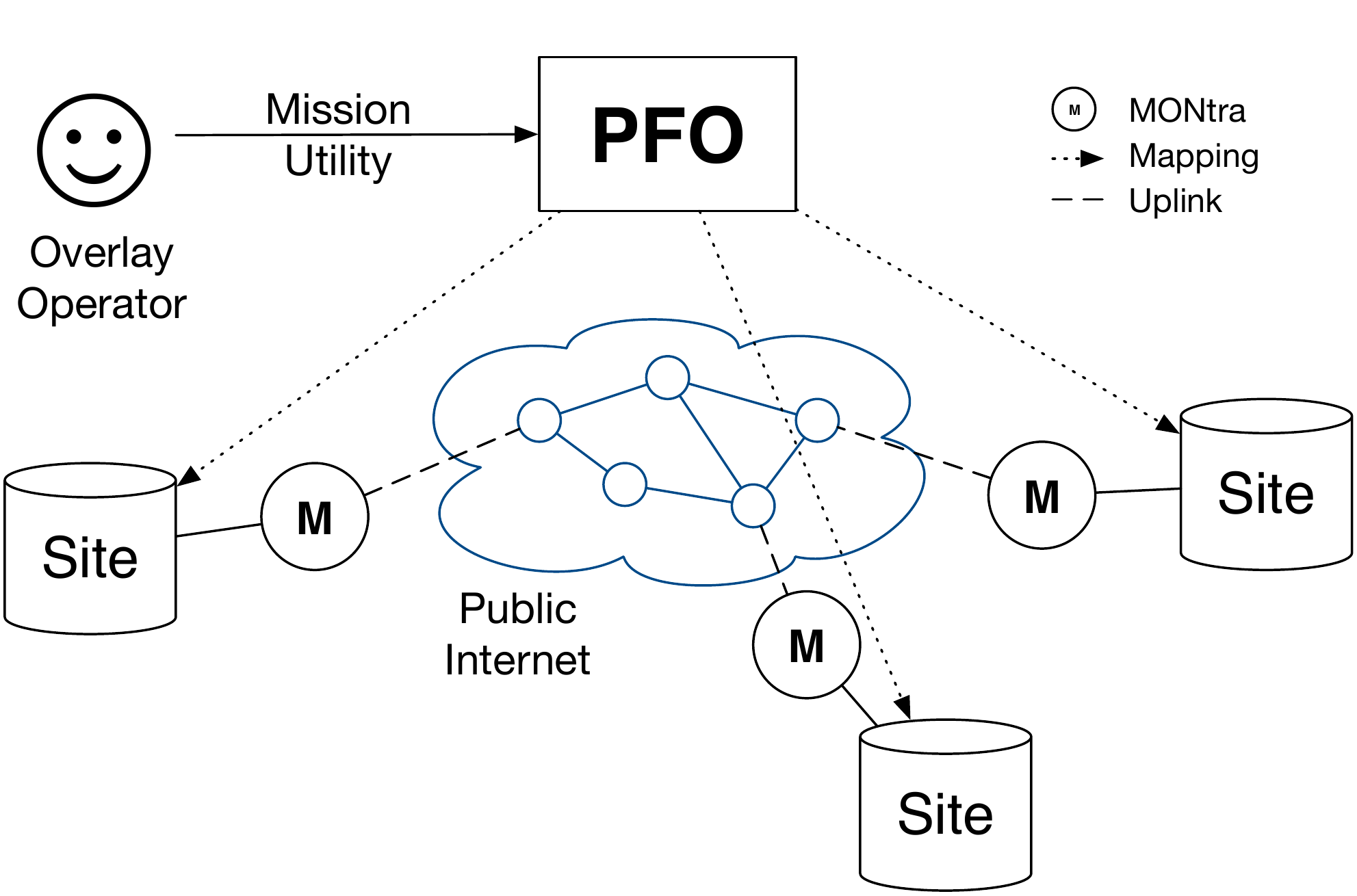}
  \caption{MON Architecture Diagram}
  \label{fig:architecture-diagram}
\end{figure}

MON takes as input the (time-varying) mission goals set by the operator and routes traffic on the overlay network to meet these goals (see Figure~\ref{fig:architecture-diagram}).  Each site is connected to the public Internet through a transport controller which performs overlay routing.

We group the traffic between sites into a set of classes $\allClasses$, where each traffic class represents a set of end-user sessions of a specific type (such as video, downloads, VOIP, etc\ldots) between a specific source and destination site. Each traffic class may use a set of overlay routes. The mission goals are captured by mission utility functions specified by the overlay operator. MON determines a set of sessions from each class and a rate for each session so as to maximize the cumulative mission utility.

% We group the traffic between sites into a set of classes $\allClasses$, where each traffic class $\class \in \allClasses$ represents a set of flows of a specific type (such as video, downloads, VOIP, etc) between a specific source and destination site. Let each class have a set of overlay routes $\rho_\class$ that it can use. Let the mission goals be captured by a mission utility function specified by the overlay operator for each class $\class \in \allClasses$. Specifically, we are given functions $U_\class(\classRate)$ that equals the incremental utility achieved for routing class $\class$ with an aggregate rate of $\classRate$ over all routes $\rho_\class$. MON chooses a number of flows $n_\class$ for each class $\class \in \allClasses$ and routes each chosen flow of class $\class$ at a rate of $\classRate$ in manner that maximizes the cumulative mission utility expressed as $\sum_\class n_\class U_\class(\classRate)$.

MON is designed to continually adapt to change. The mission utility functions can change as mission goals change. The number of sessions in each class that need to be routed can change with user demand. The underlying Internet could suffer from failures that require traffic to be rerouted. To deal with these changes,  MON continually adapts the number of sessions $n_\class$ and rates $\classRate$ for each class $\class \in \allClasses$ to optimize the mission utility.

\subsection{Our Contributions}

We propose a novel two-tiered overlay architecture for MON that combines offline and online tiers. The offline tier is called the Predictive Flow Optimizer (PFO), and it periodically performs a global optimization of the cumulative mission utility. The output of PFO is ``mapped'' to a lower-level online network transport mechanism called MONtra which performs the actual routing of traffic in the network using proportionally-fair utility functions (see Figure~\ref{fig:architecture-diagram}). We prove that MON's two-tiered architecture converges to an optimal cumulative mission utility. An interesting aspect of our work is a mapping process that allows us to implement arbitrary non-decreasing mission utility functions using logarithmic transport utilities which are well-studied and have desirable properties such as proportional fairness.

To establish the real-world feasibility of MON, we implement a prototype within the Deterlab \cite{Mirkovic10thedeter} testbed. We show that the PFO implementation using a bilinear global optimizer, in combination with MONtra implemented on the Deterlab nodes, is able to send traffic at rates that converge to a solution that achieves optimal mission utility. We also show that the system is robust to changes in the network, such as those caused by network partitions or congestion. Further, we show that the system is robust to changes in the number of sessions, such as those caused by flash crowd events. We also empirically evaluate MON when the network and traffic demands are not precisely known. In this case, we show that MON degrades gracefully and still provides a near-optimal mission utility.

%Specifically, PFO uses the current mission utilities and predictions for the traffic demands and the %predicted state of the network to compute the following for each class $c$:  a feasible number of flows %$n_c$,  the desired rate $\classRate$ for each admissible flow, and the overlay route $\rho_c$ from source to %desitnation. Note that PFO performs both flow admission and flow routing to maximize the cumulative %mission utility.  The PFO flow admission and routing decisions are then ``mapped'' to the MONTra %nodes (see Figure~\ref{fig:architecture-diagram}) that uses logarithmic utility funtions to route the flows %in a proportionally-fair manner, setting rates based on the {\em actual} real-time state of the Internet.

\subsection{Roadmap}

We give an overview of the MON architecture and a detailed description of each component in Section \ref{section:approach}. We describe our prototype and experimental setup in detail in Section \ref{section:methodology}. We implement MON and present our empirical results in Section \ref{section:results}. We compare MON to prior work in Section \ref{section:related} and then conclude in Section \ref{section:conclusion}.

\section{The MON Architecture} \label{section:approach}

MON dynamically allocates available overlay resources to traffic between sites to maximize overall mission utility. In order to do this, we need a precise definition of the mission utility maximization problem. We group the traffic between sites into a set of classes $\allClasses$, where each traffic class $\class \in \allClasses$ represents a set of sessions of a specific type (e.g. video or VOIP) between a specific source and destination site. Each class uses a set of overlay routes $\rho_\class$. Mission goals are captured by mission utility functions specified by the overlay operator. Specifically, associated with each class $\class \in \allClasses$ is a function $U_\class(\classRate)$ that corresponds to the value of one session of class $\class$ receiving a rate of $\classRate$. MON chooses a number of sessions $n_\class$ for each class $\class \in \allClasses$ and routes each chosen session of class $\class$ at a rate of $\classRate$ so as to maximize the cumulative mission utility expressed as $\sum_\class n_\class U_\class(\classRate)$.

MON has a two-tiered architecture which combines a non-real-time global optimizer (PFO) with a distributed real-time transport protocol (MONtra) to optimize mission utility.  It is a novel application of the divide-and-conquer principle in network design. We use predicted global knowledge to periodically ``push'' the overlay network into an optimized state. We maintain the network in a near-optimal state, even in the presence of sudden network changes (such as partitions or congestion), using a mission-aware distributed transport protocol. In this section, we describe the following three major aspects of the architecture (see Figure~\ref{fig:architecture-diagram}):
\begin{itemize}
\item The Predictive Flow Optimizer (Section \ref{section:PFO}) solves an  optimization problem to come up with a plan for routing traffic in the MON. It solves a non-concave bilinear optimization problem periodically using the predicted network state and projected future traffic conditions.
\item MONtra (Section \ref{section:montra}) solves an online optimization problem to react to changes in the network. Using ideas from network utility maximization \cite{Kelly97chargingand}, it adjusts the sending rates of each site to solve a convex optimization problem.
\item A mapping between PFO and MONtra (Section \ref{section:mapping}) ensures that when PFO has full knowledge of the network, MONtra will converge to PFO's target rates. Our main result is that this convergence happens if MONtra has the same gradient as PFO at the target rates.
\end{itemize}

\subsection{Predictive Flow Optimizer} \label{section:PFO}

\begin{figure}
  \centering
  \begin{tabular}{|r|l|}\hline%
    \bfseries Symbol & \bfseries Meaning \\ \hline
    $\allClasses$ & Set of traffic classes  \\ \hline
    $N_k$ & Maximum number of sessions for class $k$  \\ \hline
    $\rho_k$ & Set of routes usable by $k$  \\ \hline
    $L$ & Set of underlay links  \\ \hline
    $\hat{L}$ & PFO's estimate of $L$  \\ \hline
    $C$ & Set of underlay link capacities  \\ \hline
    $\hat{C}$ & PFO's estimate of $C$  \\ \hline
    $U_k(x_k)$ & Per-class mission utility  \\ \hline
    $n_k$ & Number of admitted sessions for class $k$  \\ \hline
    $\routeRate$ & Rate assigned to a flow  \\ \hline
    $\classRate$ & Aggregate class rate ($\classRate = \sum_{\flow \in \rho_k} \routeRate$)   \\ \hline
    $w_\classRoutePair$ & MONtra's weight for class $k$ on flow $f$ \\ \hline
    $\transportUtility$ & MONtra's utility for class $k$ on flow $f$ \\ \hline
    $\gamma$ & MONtra's stability constant \\ \hline
  \end{tabular}
  \caption{Table of Notation}
  \label{table:notation}
\end{figure}

The Predictive Flow Optimizer (PFO) outputs a routing plan for the network that maximizes mission utility. It runs periodically using a prediction of future network conditions and traffic demands.

PFO performs ``call admission'' by choosing a number of sessions $n_\class$ to admit in each class $\class \in \allClasses$. It can decide to admit no sessions at all for a given class $\class$ by setting $n_\class$ to zero. In addition, PFO chooses a per-flow rate $\routeRate$ to provide to each admitted session of class $\class \in \allClasses$ along a route $\flow \in \rho_\class$ in the network. The output of PFO is then used to set the parameters of the MONtra controllers, a process we call ``mapping''. Thus, PFO solves a hard global optimization problem, albeit in a non-realtime fashion using predicted traffic and network states.

{\bf The Optimization problem:} PFO runs periodically and solves the following optimization problem to route a predicted set of sessions on the overlay. PFO takes as input a set of traffic classes $\allClasses$.  Each traffic class $\class \in \allClasses$ has a set of $N_\class$ sessions that need to be routed from a specific source site to a specific destination site. For instance, a traffic class could be all the VOIP phone calls made from a given site to another given site.

PFO has the option to send traffic along different paths in the network. For example, it might send traffic directly from one site to another, or send it indirectly via a number of enclaves. We say that a {\it flow} corresponds to the unique pair of a traffic class and a route through the network. For each class $\class$, PFO has a set of $\rho_\class$ of possible flows. Each flow $f \in \rho_\class$ starts at the source and ends at the destination associated with the class, using zero or more sites as intermediate nodes. Let $\rho$ denote the set of flows for all classes.

% Let $\rho = \bigcup_{\class \in \allClasses} \rho_\class$ be the set of all overlay routes.

PFO uses a model of the underlying network to pick a feasible set of rates for the flows in each class. Let $L$ be the set of underlay network links, and $C$ the set of link capacities. The capacity of link $l \in L$ is $C_l$. For convenience, we will write $l \in f$ to denote the links used by flow $f$ and $l \ni f$ to denote the flows that use link $l$. PFO uses an estimate of the set of links $\hat{L}$, and an estimate $\hat{C}$ of the link capacities.

PFO is said to have {\em full knowledge} of the network if $L=\hat{L}$ and $C = \hat{C}$. At a minimum, PFO knows the uplinks for each site and their capacities, i.e., $\hat{L}$ is the set of uplinks from the MONtra nodes to the public Internet (see Figure~\ref{fig:architecture-diagram}). Between these two extremes, PFO may incorporate partial knowledge of the links and capacities using tools from network tomography (e.g. \cite{castro2004network}).

% PFO uses a $(|C| \times |\rho|) \times |L|$ routing matrix $\hat{R}$,
% where $\hat{R}_{(c,r),l} = 1$ if class $c$ and route $r \in \rho$ uses link $l \in L$, and is $0$ otherwise.

% PFO also estimates the capacity vector $\hat{C}$, where $\hat{C_l}$ is the bandwidth of link $l \in L$.

In addition to the above, the overlay operator provides PFO a mission utility function $U_\class(\classRate)$, for each traffic class $\class$, representing the value of giving one session of class $\class$ a rate of $\classRate$. We assume that the mission utility functions are from $\mathbb{R}^+ \rightarrow \mathbb{R}$, are subdifferentiable everywhere, and are non-decreasing. We do {\it not} assume that they are concave, in order to incorporate mission utilities for inelastic traffic \cite{1582432}.

For each traffic class $\class \in \allClasses$, PFO outputs a number of allowed sessions $n_\class$. For each class $\class$ and possible flow $f \in \rho_\class$, PFO outputs a target rate $\routeRate$ which corresponds to the amount of traffic MON sends for a {\it single} session along flow $f$. To do so, PFO solves the following non-concave optimization problem:
\begin{align}
    \underset{n,x}{\text{max }} & \sum_{\class \in \allClasses} n_{\class}U_\class(\sum_{\flow \in \rho_\class}\routeRate) \label{eq:pfo-optimization-problem} \\
    \text{subject to }
    % & \sum_{r \in \rho_\class}\routeRate = \classRate  & \forall \class \in \allClasses  \\
                                & \sum_{\class \in \allClasses} \sum_{\flow \in \rho_\class; l \ni \flow} n_{\class} \routeRate \leq \hat{C_l} & \forall l \in \hat{L} \label{eq:pfo-rate-constraint} \\
    % & \hat{R}x \leq \hat{C} \\
    & n_\class \leq N_\class & \forall \class \in \allClasses \nonumber \\
    & \routeRate \geq 0 & \forall \class \in \allClasses, \flow \in \rho_\class \nonumber \\
    & n_\class \in \mathbb{Z} & \forall \class \in \allClasses \nonumber
\end{align}

{\bf Solving the Optimization Problem:}
The above optimization problem is NP-Hard, since the number of sessions must be an integer and the mission utility functions are not concave. However, for certain mission utility functions there are optimization techniques which make solving this problem more feasible. For instance, if the mission utility functions are piecewise linear, the problem becomes a bilinear program that can be solved efficiently using the ANTIGONE solver \cite{misener-floudas:ANTIGONE:2014}. Other approaches for solving non-concave network utility maximization problems offline are described in the literature (e.g. \cite{1582432}).

\subsection{MON Transport Control (MONtra)} \label{section:montra}

MONtra works at the transport layer of MON, and is responsible for reacting rapidly  to changes in the underlying network. MONtra consists of weighted proportionally-fair congestion controllers, which route session traffic to match the rate chosen by PFO.

For each overlay route, MONtra's controllers optimize the transport-layer utility function $\transportUtility = w_\classRoutePair\log{\routeRate}$ (which should not be confused with the mission utility function $U_\class(x)$). In Section \ref{section:mapping}, we will describe how to choose weights for these controllers so that they provably converge to PFO's target rates. In an attempt to make the mapping easier to understand, we will model MONtra as using one controller per session on a flow. It's possible to extend the model to combine all the sessions on a flow into one controller. MONtra solves the following optimization problem, where $L$ is the set of links in the network and $C_l$ is the capacity of a link:
\begin{equation*}
  \begin{aligned}
    \underset{x}{\text{max }} & \sum_{\flow \in \rho} n_{\class(\flow)} V_\classRoutePair(x_\classRoutePair) \\
    \text{subject to }
    & \sum_{\class \in \allClasses} \sum_{\flow \in \rho_\class; l \ni \flow} n_{\class} \routeRate \leq C_l & \forall l \in L \\
    & \routeRate \geq 0 & \forall \class \in \allClasses, \flow \in \rho_\class
  \end{aligned}
\end{equation*}

We solve this optimization problem using techniques from Network Utility Maximization \cite{Kelly:2005:SEA:1064413.1064415}. We initialize the controllers to the the rate selected by PFO, then adapt the rate based on network feedback. After each success/loss signal, we adjust rates according to the following update rules, where $\gamma$ is a constant chosen for stability:
\begin{align*}
  \routeRate &\leftarrow \routeRate + \gamma \cdot w_\classRoutePair & \text{ (after each successful packet)} \\
             &\leftarrow \routeRate - \gamma \cdot \routeRate  & \text{ (for each loss)}
\end{align*}

Note that unlike existing multipath TCP research (e.g. \cite{Kelly:2005:SEA:1064413.1064415, wischik2011design, khalili2012mptcp, han2006multi, peng2016multipath}), MONtra uses uncoupled controllers. We would like our controllers to exactly match the target rates chosen by PFO, which implies the transport optimization problem should have a unique optima. Unfortunately, the coupled controllers in the multipath literature allow multiple optima.

Instead of using a window-based controller, we use a rate based controller which sends packets at a rate of $\routeRate$. To do this, we generate delays between each packet so that the packet sending process is Poisson with rate $\routeRate$.

\subsection{Mapping} \label{section:mapping}

The mapping layer is responsible for ensuring that MONtra converges to the set of rates chosen by PFO. Intuitively, we would like the transport utility function to act like the mission utility functions in the vicinity of the target rates selected by PFO. If we could ensure that MONtra sends at the same rate as PFO {\it and} has the same derivative at PFO's target rates, MONtra might behave in the same way as PFO even if there were slight changes to the network. The following theorem uses similar intuition and allows us to prove that MONtra converges to PFO's target rates.

% The core of the proof is the following lemma:

% \begin{restatable}{lemma}{mappinglemma} \label{lemma:mapping}
%   Let $f(x)$ be any differentiable function and let $c(x)$ be a differentiable convex function. Let $A$ be a minimum of $f(x)$ over a convex feasible region, and let $A'$ be a minimum of $c(x)$ over the same feasible region. If $\nabla c(A) = \nabla f(A)$, then $A' = A$.
% \end{restatable}

% \begin{IEEEproof}
% See Appendix \ref{appendix:mapping-proof}.
% \end{IEEEproof}

\begin{restatable}{theorem}{mappingtheorem} \label{theorem:mapping-works}
  Suppose PFO has full knowledge of the network and selects a set of rates $A$ and a number of sessions $n_\class$ for each class $\class$ that maximizes the mission utility function $U(A)$. For each link $l$, let $\lambda_l$ be the dual variable associated with the capacity constraint for link $l$. Fix the number of sessions for each class in the transport layer to $n_\class$. Using the following transport utility functions for a given flow $f$ with class $\class$, MONtra's rates will converge to $A$:
\begin{equation*}
\begin{aligned}
  \transportUtility &= w_\flow \log{\routeRate} \\
  w_\flow &= n_\class \big(\sum_{l \in \flow} \lambda_l \big)A_\flow
\end{aligned}
\end{equation*}
\end{restatable}

\begin{IEEEproof}
See Appendix \ref{appendix:mapping-proof}
\end{IEEEproof}

Note that if the only active constraint in the PFO solution is constraint \eqref{eq:pfo-rate-constraint}, this mapping confirms the earlier intuition that we should match the PFO gradient at the target operating point. Since only the rate-related constraints are active for PFO, by the PFO KKT conditions, $\frac{\partial}{\partial \routeRate} U(A) = (\sum_{l \in \flow} n_\class\lambda_l)$. Therefore, our mapping simplifies to $w_\flow = \frac{\partial}{\partial \routeRate} U(A) A_\flow$. At the target operating point, the partial derivative of the MONtra utility function with respect to a flow $\flow$ is $\frac{w_\flow}{A_\flow}  = \frac{\partial}{\partial \routeRate} U(A)$. So in addition to matching the rate, in this case we would also expect the MONtra utility functions to approximate the PFO utility functions close to the operating point.

This mapping theorem also works for any implementation of MONtra and other formulations of PFO optimization problem. For instance, MONtra could use other classes of concave transport utility functions, or another method of distributed optimization such as backpressure routing. The PFO optimization problem could use another way of combining per-flow rates, e.g. by summing the {\it weighted} rates across all paths instead of summing the rates across all paths.

% \subsubsection{fixed packet loss assumption}

% Suppose we assume that PFO has full knowledge of the network and that the amount of traffic we are sending our traffic is small enough that it doesn't affect packet loss rates.

% Let $p$ be the packet loss rate and $k$ be a constant chosen for stability. The behavior of MONtra can then be described by the following differential equation (from \cite{Srikant:2014:CNO:2636796}, Section 2.4):

% $$ \dot{\routeRate} = k\routeRate(\frac{w_\classRoutePair}{\routeRate} - p) $$

% At equilibrium, $\dot{\routeRate} = 0$ so the rates on each path will satisfy:

% $$ 0 = \frac{w_\classRoutePair}{\routeRate} - p $$

% Since we've assumed $p$ is a constant, we can set $w_\classRoutePair$ so that the flow will converge to PFO's output $\targetRouteRate$:

% $$ w_\classRoutePair = \targetRouteRatep $$

% \subsubsection{matching value and derivative}

% This mapping is not ideal, since our controllers could affect packet loss.

%%% Local Variables:
%%% mode: latex
%%% TeX-master: "paper.tex"
%%% End:

\section{Evaluation Methodology} \label{section:methodology}

\begin{figure}
  \centering
  \subfloat[Triangle Topology]{
    \includegraphics[width=0.15\textwidth,keepaspectratio=true]{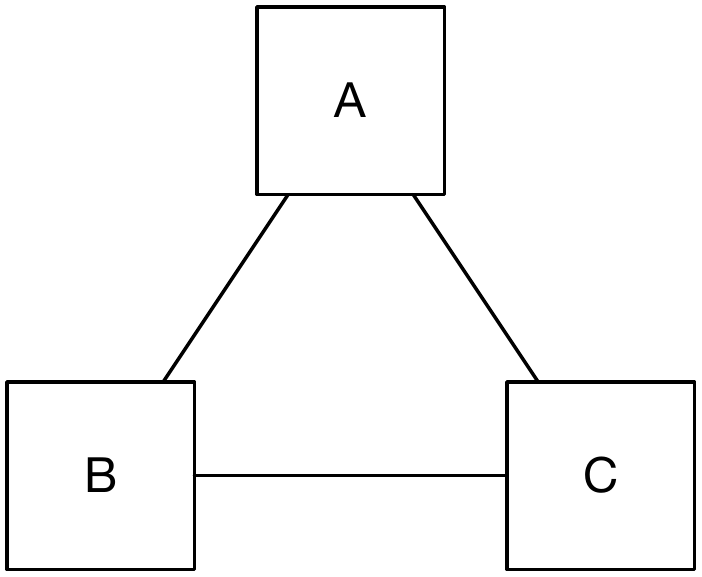}
    \label{fig:topology:triangle}
  }
  \qquad
  \subfloat[AT\&T Topology]{
    \includegraphics[width=0.45\textwidth,keepaspectratio=true]{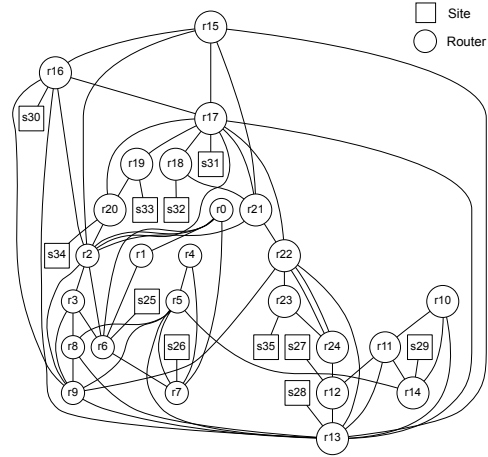}
    \label{fig:topology:att}
  }
  \caption{Experimental Topologies}
  \label{fig:topology}
\end{figure}
To show how MON performs in a realistic setting, we implemented MON and ran it on a set of network topologies, traffic scenarios, and mission utility functions as outlined below.

{\bf Network Testbed.} We ran the experiments on Deterlab \cite{Mirkovic10thedeter}, which allowed us to allocate physical linux machines for each site and router in the network. We used Linux's traffic control system to set the network bandwidth. We used token bucket filters with a burst size of $100kb$ and a maximum queue latency of $5ms$, which provides stable throughput when we transfer files between the hosts.

{\bf Network Topologies.} We emulated the small triangle topology shown in Figure \subref*{fig:topology:triangle} to illustrate MON behavior in an easier to understand context. We also emulated several large topologies from \cite{topology-zoo} (AT\&T USA, Bell Canada, BTN, and Abilene). We present the results for the AT\&T topology shown in Figure~\subref*{fig:topology:att} in most of our experiments.

{\bf Mission Utility Functions.} We use the following two mission utility functions to illustrate the behavior of MON, though our system works for arbitrary mission utility functions:
\begin{align*}
    U_A(x) &=
\begin{cases}
    0 & \text{if } x \leq 0.8\\
    \min(0.1x, 0.005x+0.114) & \text{otherwise}
\end{cases} \\
U_B(x) &= 0.2x
\end{align*}

\begin{figure}
  \centering
  \subfloat[Class A]{
    \includegraphics[width=0.2\textwidth,keepaspectratio=true]{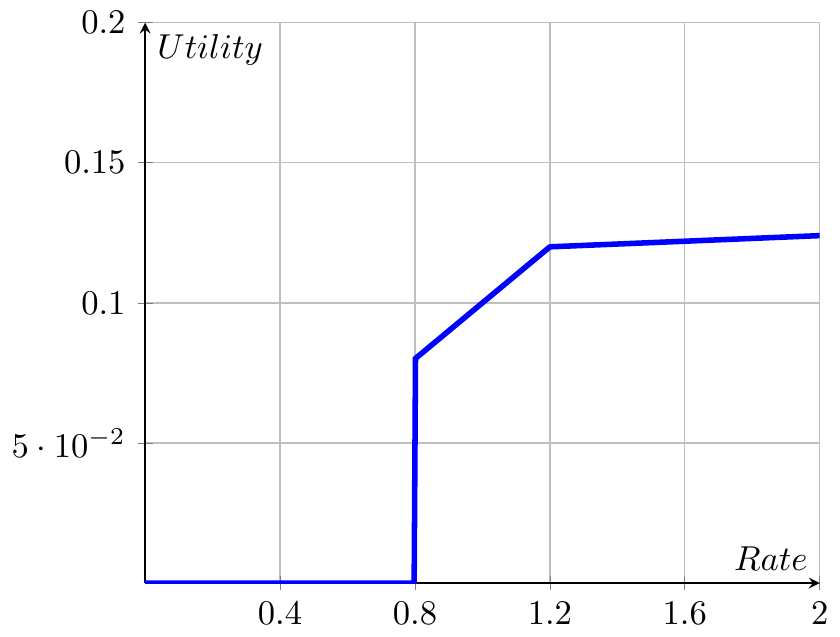}
     }
  \qquad
  \subfloat[Class B]{
    \includegraphics[width=0.2\textwidth, keepaspectratio=true]{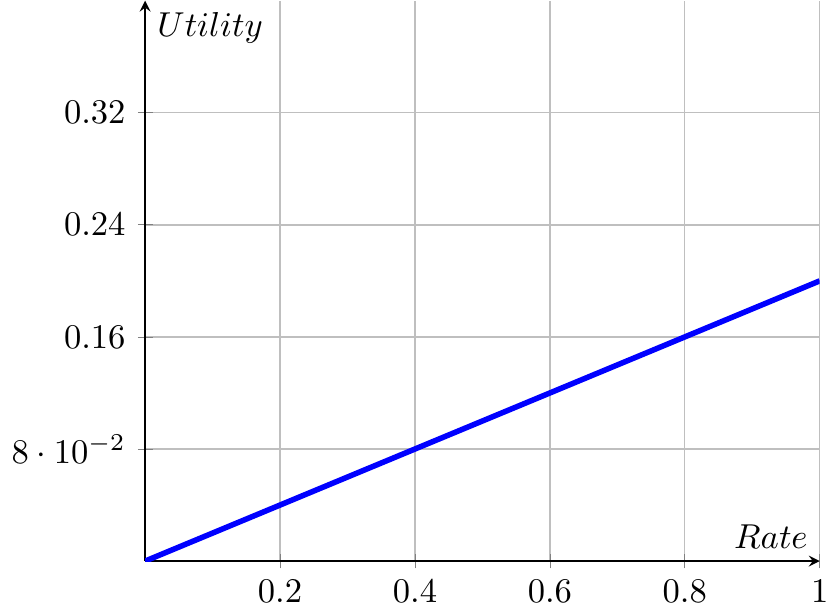}

  }
  \caption{Mission Utility Functions for Traffic Class A and B}
  \label{fig:empirical:utility functions}
\end{figure}

The type $A$ mission utility function has non-concavity and monotonically increasing utility with diminishing returns. The type $B$ mission utility function increases linearly with rate without a point of diminishing returns. These functions are shown in Figure \ref{fig:empirical:utility functions}.

\begin{figure}
  \centering
\end{figure}

{\bf PFO Implementation.} Since our approach reduces the PFO optimization to a bilinear program, we used the ANTIGONE solver \cite{misener-floudas:ANTIGONE:2014} to efficiently solve the optimization problem. The solver uses branch-and-bound techniques and convex relaxations such as McCormick's envelopes which allow bilinear optimization problems to be solved efficiently.

{\bf MONtra Implementation.} The MONtra implementation is based on multipath network utility maximization theory from \cite{Kelly:2005:SEA:1064413.1064415}. We used the controllers described in Section~\ref{section:montra} with per-packet acks to detect congestion. We gave each host an infinite backlog of data to send. We always used the number of sessions chosen by PFO. We set the stability constant $\gamma = 0.001$ and further improved stability by dividing $\gamma$ by the largest weight (i.e. $\max_{r \in \rho} w_r$).

%%% Local Variables:
%%% mode: latex
%%% TeX-master: "paper.tex"
%%% End:

\section{Evaluation Results} \label{section:results}

% Show that it works
% x do PFO and MONtra agree?
%   - triangle network, AT\&T network
% x what if the network changes a bit?
%   - triangle network - vary capacity on a link, step size 2mbps, from 2 - 20?
%   - AT\&T network - cross-network partition
% - what if the network topology isn't what PFO expects?
% x does it work on a real network? AT\&T, BellCanada
% - how badly does it work with background traffic?
% - does overlay network help at all?
% - number of overlay flows?
% - does PFO allocate more rate to flows with higher priority?

\subsection{Does the overlay optimize mission utility?}

Our first experiment shows that MON optimizes mission utility for simple scenarios. We give PFO full knowledge of the network topology and capacities, and show that MONtra converges to PFO's target rates. We show this for both a simple topology and a more realistic one.

For the simple topology, we set up three nodes on Deterlab using the triangle topology shown in Figure \subref*{fig:topology:triangle}. We set the capacity between Node B and Node C to 5Mbps and set the capacity of all other links to 10Mbps. We have two traffic classes, one between Node A and Node C and one between Node B and Node C. PFO assigns a rate of 10 Mbps between Node A and Node C and a rate of 5 Mbps from Node A to Node C via Node B. Table \subref*{table:triangle-rates} shows the rates MONtra converged to for the two flows. In this simple example, there are no shared links between the flow. Note that the actual rates achieved by MONtra is close to the target rate set by PFO.

We also ran this experiment on the more realistic AT\&T network topology. We set the capacity of all links to 10Mbps, and ran PFO using the actual topology and capacities. Unlike the triangle experiment, PFO shared links between flows. Table \subref*{fig:topology:att} shows the rates MONtra converged to, which are close to the rates computed by PFO.

% \begin{figure}
%   \centering
%   \subfloat[]{
%   \includegraphics[width=0.45\textwidth,keepaspectratio=true]{plots/implementation/basic/goodput.png}
%   \label{fig:implementation:basic:triangle}
% }
%   \qquad
%   \subfloat[]{
%   \includegraphics[width=0.45\textwidth,keepaspectratio=true]{plots/implementation/att/goodput.png}
%   \label{fig:implementation:basic:att}
% }
%   \caption{With full knowledge of the network, MONtra converged to the same rates that PFO picked for both the triangle topology \ref{fig:implementation:basic:triangle} and the AT\&T topology \ref{fig:implementation:basic:att}}
% \end{figure}

\begin{figure}
  \centering
  \subfloat[]{
    \label{table:triangle-rates}
    \begin{tabular}{l|c|c}\hline%
      \bfseries Path & \bfseries Target  & \bfseries Actual
                                           \begin{filecontents*}{basic-results.csv}
path,target,predicted,actual
$A \rightarrow C$,10.0,10.0,9.71
$A \rightarrow B \rightarrow C$,5.0,5.0,4.84
\end{filecontents*}
\csvreader[head to column names]{basic-results.csv}{}%
                                           {\\\hline \path & \target & \actual}%
    \end{tabular}
  }
  \qquad
  \subfloat[]{
    \begin{tabular}{l|c|c}\hline%
      \bfseries Path & \bfseries Target  & \bfseries Actual
                                                                  \begin{filecontents*}{att-results.csv}
path,target,predicted,actual
$s35 \rightarrow s26$,10.0,10.0,9.24
$s29 \rightarrow s31$,10.0,10.0,9.24
$s33 \rightarrow s35$,10.0,10.0,9.24
$s31 \rightarrow s33$,10.0,10.0,9.23
$s34 \rightarrow s32$,6.0,6.0,5.9
$s26 \rightarrow s34$,6.0,6.0,5.68
$s32 \rightarrow s29$,6.0,6.0,4.4
$s32 \rightarrow s34$,4.0,4.0,3.55
$s26 \rightarrow s32$,4.0,4.0,3.4
$s34 \rightarrow s29$,4.0,4.0,3.24
\end{filecontents*}
\csvreader[head to column names]{att-results.csv}{}%
                                                                  {\\\hline \path & \target & \actual}%
    \end{tabular}
  }
  \caption{MONtra rates converged to the optimal target rates set by PFO for both the triangle topology  and the AT\&T topology}
\end{figure}

% To bring in more realism into the experiments, the AT\&T network as shown in figure y was emulated on deter lab. The AT\&T network topology was obtained from [reference]. A few of the routers were assigned to be POP's at random. The reference also provided the latitude and longitude information of each of the routers. Thus the route between any two POP's was found by finding the shortest distance between the two POP's. Flows of two types as shown in figure y with the utility functions mentioned above were generated between the POP's. Apart from a direct path between the flows, alternate two hop paths were also considered. PFO was run to find the number of connections and the rate to be allocated on each path for each flow. MONtra uses the rates that PFO provides based on our formulation. It is observed that the rates achieved by MONtra is equal to rates allocated by PFO for each of the flows. \todo{Add the plots here and explain each of the plots}

\subsection{Are overlay paths useful? If yes, how many? How does random path selection compare with choosing the best paths?}

\begin{figure}
  \centering
  \resizebox{0.45\textwidth}{!}{%
    \input{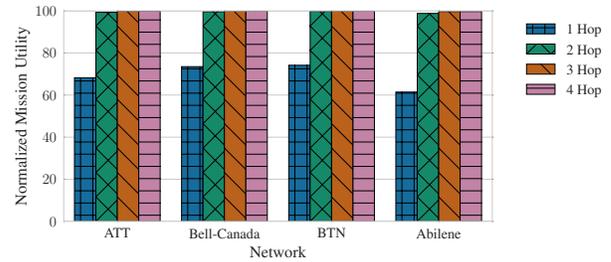}
  }
  \caption{Mission utility increases with the number of allowable hops, though two hops are sufficient to obtain most of the benefits.}
  \label{fig:empirical:benefits-of-overlay-paths}
\end{figure}

\begin{figure}
  \centering
  \resizebox{0.45\textwidth}{!}{%
    \input{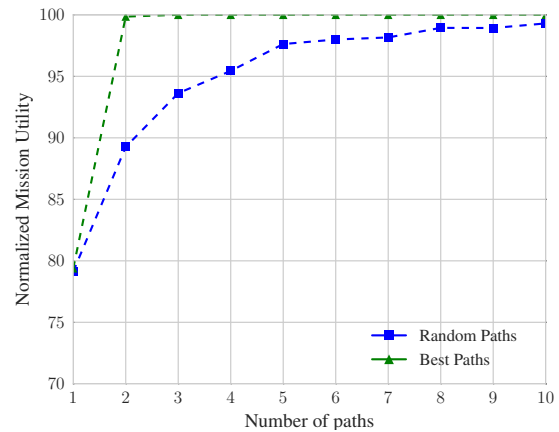}
  }
  \caption{The impact of using additional indirect paths on mission utility.}
  \label{fig:empirical:number-of-overlay-paths}
\end{figure}

Two sites can communicate directly with each other, or they can communicate indirectly via a series of other sites. The impact of overlay routing on reliability and performance for traditional best-effort overlays are well-known  \cite{AndersenSnoerenBalakrishnan2003,AndersenBalakrishnanKaashoekEtAl2001,RahulKSB06}.
Here we ask analogous questions for mission-optimized overlays by ascertaining the benefits of overlay paths for mission utility maximization.

To study the impact of overlay paths, we ran PFO on a variety of real network topologies from \cite{topology-zoo} (AT\&T USA, Bell Canada, BTN, and Abilene). We set uplink capacity to 30Mbps and the capacity between routers to 10Mbps. To simulate a partially loaded overlay, we created traffic classes between half the sites. We first restricted PFO to use only direct, one-hop paths, then to using one-hop and two-hop paths, and then to three- and four-hop paths. Note that the optimal mission utility cannot decrease when we allow a greater number of hops, since more hops corresponds to a larger feasible region. Figure \ref{fig:empirical:benefits-of-overlay-paths} shows the results of this experiment. In all networks, there was more than a 20\% increase in mission utility from using two hop paths over just the one-hop direct path. In this particular experiment, there was no benefit to using paths longer than two hops.

Next, we consider the impact of the number of allowed paths (i.e. $|\rho_\class|$) on mission utility. It may be useful to not have to consider all overlay paths when running PFO. As shown in Figure~\ref{fig:empirical:number-of-overlay-paths}, adding just one well-chosen indirect path for each class can be sufficient to obtain the maximum mission utility. However, we generally do not know the single best path before running PFO. Including a few random paths is also sufficient to improve mission utility. Just one randomly picked indirect path gave 80\% of the optimal mission utility, and adding four random indirect paths gave 95\% of optimal.

% \todo{any reference which shows half the data centers are inactive or atleast some percentage of them are, otherwise the result is looks obvious}.It was observed that there is atleast a 20\% increase in the overall utility of the network by considering two hop paths in addition to direct paths. But considering more than two hop paths gives marginal increase in utility. It was also observed that most of the rate allocated is allocated on the direct and one hop paths. This can be observed in fig y.
% \todo{Add how the experiment was performed}

% \begin{figure}
%   \centering
%   \resizebox{0.45\textwidth}{!}{%
%     \input{plots/num_routes.pgf}
%   }
%   \caption{Num routes}
%   \label{fig:empirical:num_routes}
% \end{figure}

\subsection{How does MONtra react to slight changes?}

\begin{figure}
  \centering
  \resizebox{0.45\textwidth}{!}{%
    \input{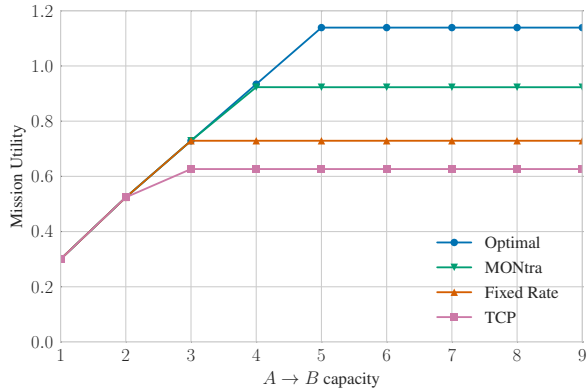}%
    \label{fig:implementation:robustness}%
  }
  \caption{How MONtra reacts when PFO has incorrect topology knowledge. We ran MONtra with PFO outputs for an $A \rightarrow C$ capacity of 3 Mbps, and plotted the rates it converged to for various values of $A \rightarrow C$ capacities.}
  \label{fig:implementation:robustness}
\end{figure}

Our next experiment shows that MONtra optimizes mission utility better than existing transport layer implementations, even when the network is slightly different from what PFO expects. We compare MONtra to a transport layer which uses Scalable TCP \cite{Kelly:2003:STI:956981.956989} without trying to match PFO's rates, and a transport layer which always sends at PFO's target rates.

This experiment used a mathematical model of the triangle network. We consider flows over links $A \rightarrow B$ and $B \rightarrow C$. We set link $B \rightarrow C$ to 5Mbps, and run PFO with the capacity on $A \rightarrow B$ set to 3Mbps. We then used those PFO rates for different capacities on $A \rightarrow B$. We tested capacities of 1Mbps to 10Mbps in intervals of 1Mbps.

Figure \ref{fig:implementation:robustness} shows the result of this experiment. When the capacity is 3Mbps and PFO has correct knowledge of the network, MONtra and the fixed-rate implementation match PFO's target rates. When link $A \rightarrow B$'s capacity is between 1 and 4 Mbps, MONtra was also able to match PFO's rates. Although MONtra had different rates between 5 and 10Mbps, it had a higher mission utility than the other two implementations.

% In this section we show that our formulation of MONtra adapts well to slight changes in network conditions. Consider the network as shown in figure x. We fix the initial capacity of the link A->C to be 7 Mbps. We run PFO to find the target rates for each of the flows. MONtra converges to the rates allocated by PFO. We now vary the capacity of the link A->C from 0 to 10 in increments of 1 Mbps. We have calculated the utilities PFO would achieve for each of these capacities offline. We now compare the utilities MONtra achieves against the utilities that PFO would achieve when the capacity of the link is varied. It can be seen from the plot in figure \todo{add the alpha plot} that the utility MONtra achieves is very close to the utility that PFO would achieve.

\subsection{What if demand changes?}

\begin{figure}
  \centering
  \resizebox{0.45\textwidth}{!}{%
    \input{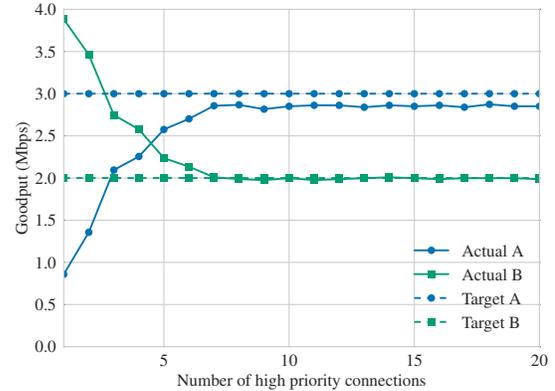}%
  }
  \caption{How MONtra reacts when demand changes. PFO chose 11 sessions for the high priority flow, but we varied the number of sessions from 1 to 20.}
  \label{fig:implementation:numflows}%
\end{figure}

In our previous experiments, we allowed PFO to choose the number of sessions used by MONtra. In a real system, the number of sessions might be less predictable; PFO may choose to admit $n_\class$ sessions and then many more sessions could arrive. In this experiment, we show how MONtra reacts when the number of sessions is more or less than what PFO chooses.

This experiment uses the triangle network. We set capacities $A \rightarrow C$ to 3Mbps, $B \rightarrow C$ to 5Mbps, and $A \rightarrow C$ to 10Mbps. PFO split the capacity of the $B \rightarrow C$ link between a flow of class A and a flow of class B. PFO gave the flow of class A 3Mbps for eleven sessions, and gave the flow of class B the remaining 2Mbps for one session.

We then vary the number of sessions for the flow of class A from 1 to 20. Figure \ref{fig:implementation:numflows} shows the rates that MONtra converged to. For 7 sessions and up, MONtra converged to PFO's target rate. Below 7 sessions, the two systems diverged and MONtra sent more of flow B than PFO.

\subsection{How does MON react to failures?}

\begin{figure*}
  \centering
  \subfloat[Rates on triangle topology]{
    \resizebox{0.45\textwidth}{!}{%
      \input{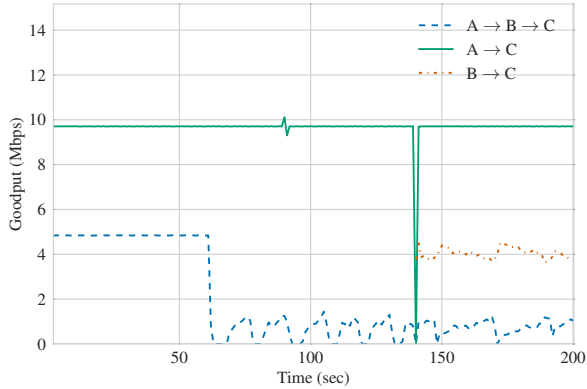}%
    }
    \label{fig:implementation:partition:triangle-rates}
  }
  \qquad
  \subfloat[Mission utility on triangle topology]{
    \resizebox{0.45\textwidth}{!}{%
      \input{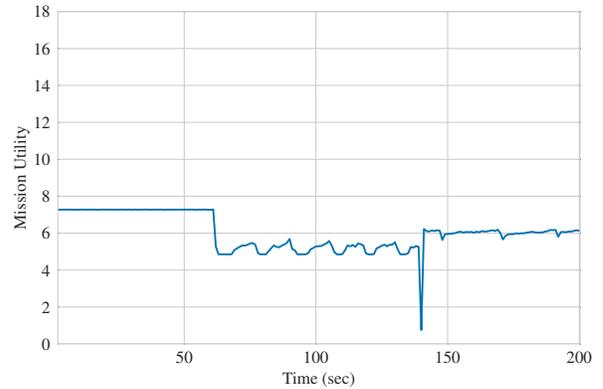}%
    }
    \label{fig:implementation:partition:triangle-utility}
  }
  \qquad
  % \subfloat[Rates on AT\&T topology]{
  % \resizebox{0.45\textwidth}{!}{%
  % \input{plots/implementation/att-partition/goodput.pgf}%
  % }
  %   \label{fig:implementation:partition:att-rates}
  % }
  %   \qquad
  %   \subfloat[Utility on AT\&T topology]{
  %   \resizebox{0.45\textwidth}{!}{%
  %   \input{plots/implementation/att-partition/total-utility.pgf}%
  % }
  %   \label{fig:implementation:partition:att-utility}%
  % }
  \caption{MONtra rates and utilities with a link failure. We started MONtra with the correct PFO outputs. At 60 seconds, we reduced the capacity of the $A \rightarrow B$ link. We allowed MONtra to adapt, then at 140 seconds, we re-ran PFO to correct the rates. The overall utility went down after the failure, though PFO was able to improve the performance over only MONtra.}
\end{figure*}

\begin{figure}
  \resizebox{0.45\textwidth}{!}{%
    \input{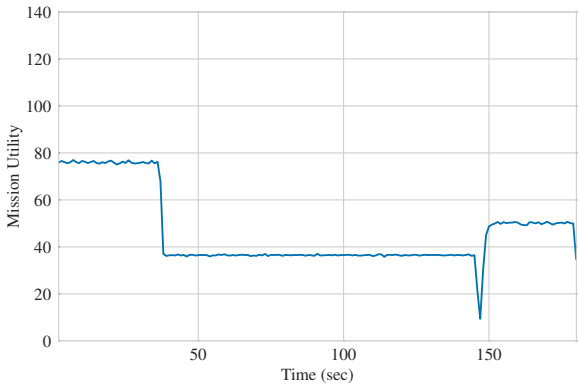}%
  }
  \caption{Mission utility over time for the partition experiment on the AT\&T network. At 40 seconds, we removed all links connected to the two most connected routers.  At 150 seconds, we re-ran PFO which increased mission utility.}
  \label{fig:implementation:partition:att-utility}%
\end{figure}

This experiment shows that MON adapts to sudden, large changes in the network. At the start of the experiment, we provide PFO full knowledge of the network and feed its output to MONtra. We then cause a link failure, and allow MONtra to react. We finally re-run PFO with the updated topology, and it picks a new, more optimal set of rates to route around the failure.

In the triangle network, we set the capacity between nodes B and C to 5Mbps, and set all other capacities to 10Mbps. MONtra begins with the rates selected by PFO. At 60 seconds, we set the capacity between nodes A and B to 1Mbps. At 140 seconds, we re-run PFO and update MONtra to use the new flows and rates. Figure \subref*{fig:implementation:partition:triangle-rates} shows the rates over time for the experiment, and Figure \subref*{fig:implementation:partition:triangle-utility} shows the mission utility over time. When the network failure occurs, MONtra adjusts its sending rate to compensate. When PFO is re-run, it selects a new set of rates that increase the overall mission utility. The drop in mission utility at 140 seconds is because our prototype doesn't gracefully switch flows after PFO runs.

In the AT\&T network, we set the capacity on all links to 10Mbps. At 40 seconds, we removed all links connected to the two routers with the highest degree ($r2$ and $r13$). Just before $150$ seconds, we re-ran PFO to select a new set of target rates. Figure \ref{fig:implementation:partition:att-utility} shows the mission utility over time. Again, MONtra reacts to the network failure, and PFO selects new target rates that increase mission utility. This shows that MON is resilient to network failures.

%%% Local Variables:
%%% mode: latex
%%% TeX-master: "paper.tex"
%%% End:

\section{Related Work} \label{section:related}

% \begin{itemize}

% % best effort overlays
% \item \cite{subramanian2004overqos} - each flow specifies the amount of packet loss they're willing to accept, and the overlay reduces bandwidth to reduce loss. can also give a certain bandwidth between nodes based on rates seen. nothing about prioritizing flows.
% \item \cite{Duan:2003:SON:965996.965998} - service overlay networks - more of a provisioning bandwidth and allocating bandwidth to links than an application utiltiy formulation
% \item \cite{Andersen:2001:RON:502034.502048} - kinda implementation-y, no congestion control (best effort)

% % NUM stuff
% \item \cite{Kelly97chargingand} - Kelly's paper - model of network utility maximization and distributed algorithm for convex problems.
% \item \todo{more distributed non-concave optimization}
% \item \cite{1582432} - Nonconcave network utility maximization - proposes an algorithm for a centralized system
% \item \cite{1664999} - tutorial on decomposition - we have stability (transport operates at faster rate than PFO)
% \item \cite{4118456} - layering as optimization decomposition - we have a vertical decomposition between pfo and transport

% % Paper from 80s in process control that has our theorem, nice parallels
% \item \cite{Biegler:1984} - plant control literature from chemical engineering has something similar to lemma 1, this was the earliest thing I could find in the literature about it
% \end{itemize}

Overlays have been studied and built for the past 25 years. Large CDNs such as Akamai \cite{akamai-overview} have built overlays for delivering web and video content since the late 1990's\cite{sitaraman2014overlay}.  However, these overlays are ``best-effort'' overlays that attempt to provide higher reliability and performance than what the native Internet can offer.  Best-effort overlays come in many flavors, including caching overlays for Web content \cite{dilley2002globally}, routing overlays for reliably transporting live video streams \cite{andreev2003designing,KontothanasisSWHKMSS04}, P2P overlays for downloads \cite{StoicaMorrisLiben-NowellEtAl2003, RatnasamyFrancisHandleyEtAl2001,ZhaoKubiatowiczJoseph2002}, and security overlays for preventing DDoS attacks \cite{sitaraman2014overlay}. However, such overlays do not attempt to explicitly optimize the ``mission goals'' of the enterprise that operates the overlay, the focus of our work.

There has been prior work on overlay networks driven by quality-of-service (QoS).  Networks that focus on QoS tend to guarantee each flow a particular performance metric such as rate, packet loss, jitter etc\cite{subramanian2004overqos, Duan:2003:SON:965996.965998}. In contrast, MON works by optimizing the cumulative mission utility of the overlay traffic. In particular, MON might sacrifice the QoS of some (lower-priority) traffic flows to enhance the QoS of other (higher-priority) flows.

MON uses the network utility maximization (NUM) framework, first pioneered by Kelly \cite{Kelly97chargingand} who described distributed algorithms for optimizing concave utilities. Non-concave NUM problems are one of the major open problems in the field (e.g. \cite{4118456}; sec. V-E). The general approaches are to solve the problem offline (e.g. \cite{1582432}), which means the system adapts slowly at best or not at all at worst to network failures, or to approximate the problem with a distributed algorithm (e.g. \cite{1498818}), which may not converge to a globally optimal point. Our two-tiered approach of periodically performing a global non-concave optimization to drive real-time transport controllers with logarithmic utilities is a novel approach to this classical problem.

% MON has some parallels in the chemical engineering literature, where a plant is controlled by the output of an optimization problem. There is a theorem in the literature similar to our Lemma 1 that gives conditions on when the combined plant and optimization system will converge to the optimization result (e.g. \cite{Biegler:1984}). However, to the best of our knowledge the result is new in computer science.

%%% Local Variables:
%%% mode: latex
%%% TeX-master: "paper.tex"
%%% End:

\section{Conclusions} \label{section:conclusion}

In this paper, we proposed a Mission-optimized Overlay Network which routes traffic to explicitly optimize the goals of an organization. By incorporating utility into its routing decisions, MON's decisions are more useful than prior, best-effort overlays. We proposed a novel, two-tiered architecture for this overlay. The higher-tier PFO periodically performs a global optimization of cumulative mission utility, allowing it to maximize the complex, non-concave utility functions that occur in practice. The lower-tiered MONtra transport protocol uses PFO's target rates to keep the overlay network near the optimal operating point by responding to network and traffic events in real-time. This architecture also opens up a new approach to investigating non-concave Network Utility Maximization (NUM) problems which has not been studied in the prior literature.

We implemented a prototype of our architecture on physical hardware, and showed that MON converges quickly to a state where the cumulative mission utility is maximized. By using mission utility information, MON can respond gracefully to network failures and changes in demand.

%%% Local Variables:
%%% mode: latex
%%% TeX-master: "paper.tex"
%%% End:

\section*{Acknowledgments}
The research reported in this paper was sponsored by DARPA under Contract No. N66001-15-C-4045. The
views and conclusions contained in this document are those of the authors and should not
be interpreted as representing the official policies of DARPA and the
the U.S. Government. The U.S. Government is authorized to reproduce and distribute reprints for Government purposes notwithstanding any copyright notation hereon. Approved for public release; unlimited distribution.

\bibliographystyle{abbrv}
\bibliography{paper,ramesh}

\appendices

\section{Proof of mapping theorem}
\label{appendix:mapping-proof}

\mappingtheorem*

\begin{IEEEproof}
  Recall the PFO solves the following optimization problem:
\begin{align*}
    \underset{n,x}{\text{max }} & \sum_{\class \in \allClasses} n_{\class}U_\class(\sum_{\flow \in \rho_\class}\routeRate) \\
    \text{subject to }
    % & \sum_{r \in \rho_\class}\routeRate = \classRate  & \forall \class \in \allClasses  \\
    & \sum_{\class \in \allClasses} \sum_{\flow \in \rho_\class; l \ni \flow} n_{\class} \routeRate \leq \hat{C_l} & \forall l \in \hat{L} \\
    % & \hat{R}x \leq \hat{C} \\
    & n_\class \leq N_\class & \forall \class \in \allClasses \\
    & \routeRate \geq 0 & \forall \class \in \allClasses, \flow \in \rho_\class \\
    & n_\class \in \mathbb{Z} & \forall \class \in \allClasses
\end{align*}

And MONtra solves the following optimization problem:
\begin{equation*}
  \begin{aligned}
    \underset{x}{\text{max }} & \sum_{\flow \in \rho} n_{\class(\flow)} V_\classRoutePair(x_\classRoutePair) \\
    \text{subject to }
    & \sum_{\class \in \allClasses} \sum_{\flow \in \rho_\class; l \ni \flow} n_{\class} \routeRate \leq C_l & \forall l \in L \\
    & \routeRate \geq 0 & \forall \class \in \allClasses, \flow \in \rho_\class
  \end{aligned}
\end{equation*}

Since we assume PFO has perfect information, $\hat{L} = L$ and $\hat{C} = C$.

Since the set of rates $A$, the number of admitted sessions $n_k$, and the dual variables for each link $\lambda_l$ and for each constraint on the number of sessions $\lambda_k$ are optimal for PFO, they satisfy the following KKT conditions (See \cite{Boyd:2004:CO:993483}; Sec. 5.5.3):
  \begin{align}
    \sum_{\class \in \allClasses} \sum_{\flow \in \rho_\class; l \ni \flow} n_{\class} \targetRouteRate &\leq C_l & \forall l \in L \label{eq:pfo-kkt-one} \\
    \targetRouteRate &\geq 0 & \forall \class \in \allClasses, \flow \in \rho_\class \label{eq:pfo-kkt-two} \\
    n_\class &\leq N_\class & \forall \class \in \allClasses \nonumber \\
    n_\class &\in \mathbb{Z} & \forall \class \in \allClasses \nonumber \\
    \lambda_l &\geq 0, & \forall l \in L \label{eq:nonzero-constraint} \\
    \lambda_l(\sum_{\class \in \allClasses} \sum_{\flow \in \rho_\class; l \ni \flow} n_{\class} \targetRouteRate - C_l) &= 0 & \forall l \in L \label{eq:slackness-condition}
\end{align}
\vspace*{-\baselineskip}
\begin{align}
  \frac{\partial}{\partial \routeRate} U(A) = \sum_{l \in \flow} n_{\class}\lambda_l \label{eq:gradient-condition} \\
  \frac{\partial}{\partial n_\class} U(A) = \sum_{k \in \class} n_{\class}\lambda_k \label{eq:gradient-condition-numbers}
  \end{align}

For a set of target rates $A'$ and link dual variables $\lambda_l'$ to be optimal for the MONtra problem, they must satisfy the following KKT conditions for the MONtra problem:
\begin{align}
  \sum_{\class \in \allClasses} \sum_{\flow \in \rho_\class; l \ni \flow} n_{\class} \targetRouteRate' &\leq C_l & \forall l \in L \label{eq:montra-kkt-one} \\
  \targetRouteRate' &\geq 0 & \forall \class \in \allClasses, \flow \in \rho_\class \label{eq:montra-kkt-two} \\
  \lambda_l' &\geq 0, & \forall l \in L \label{eq:montra-kkt-three} \\
  \lambda_l'(\sum_{\class \in \allClasses} \sum_{\flow \in \rho_\class; l \ni \flow} n_{\class} \targetRouteRate' - C_l) &= 0 & \forall l \in L \label{eq:montra-kkt-four}
\end{align}
\vspace*{-\baselineskip}
\begin{align}
  \frac{d}{d \routeRate} \transportUtility &= \sum_{l \in \flow} n_\class\lambda_l' \label{eq:montra-kkt-five}
\end{align}

  Since the MONtra problem is concave, if some point $A',\lambda_l'$ satisfy the MONtra KKT conditions then it is optimal and MONtra will converge to that point. We will show that PFO's optimal rates $A$ and link dual variables $\lambda_l$ satisfy these KKT conditions, so MONtra will converge to the PFO optimal point.

  The MONtra KKT conditions \eqref{eq:montra-kkt-one} and \eqref{eq:montra-kkt-two} are satisfied since they are identical to the PFO KKT conditions \eqref{eq:pfo-kkt-one} and \eqref{eq:pfo-kkt-two} because we've fixed the number of flows in the transport to the number of flows chosen by PFO. MONtra KKT condition \eqref{eq:montra-kkt-three} is satisfied by PFO KKT condition \eqref{eq:nonzero-constraint}. MONtra KKT condition \eqref{eq:montra-kkt-four} is satisfied by the PFO KKT condition \eqref{eq:slackness-condition}. MONtra KKT condition \eqref{eq:montra-kkt-five} is satisfied by our choice of $w_\flow$ since for flow $f$ with class $k$:
\begin{align*}
  \frac{d}{d \routeRate} \transportUtility &= \frac{w_\flow}{A_\flow} \\
                            &= \sum_{l \in \flow} n_\class\lambda_l
\end{align*}

Therefore $A,\lambda_l$ satisfy the KKT conditions for MONtra. Since MONtra is concave, it will converge to $A,\lambda_l$.
\end{IEEEproof}

\end{document}